\documentclass[journal]{IEEEtran}
\usepackage{amsmath}
\usepackage{amsfonts}
\usepackage{mathrsfs}
\usepackage{graphicx}
\usepackage{algorithm}
\usepackage{algpseudocode}
\usepackage{indentfirst}
\usepackage{enumerate}
\usepackage{subfigure}
\usepackage{epstopdf}
\usepackage{cite}
\usepackage{booktabs}
\usepackage{amssymb}
\usepackage{multicol}
\usepackage{bm}
\usepackage{subfigure}
\usepackage{caption}
\usepackage{makecell,rotating,multirow,diagbox}
\usepackage{ntheorem}
\usepackage{color}
\usepackage{ylqsymbol}
\usepackage{enumerate}
 \usepackage[normalem]{ulem}
\ifCLASSINFOpdf
\else
\fi
\begin{document}

\title{Estimating Doubly-Selective Channels for Hybrid mmWave Massive MIMO Systems:\\ A Doubly-Sparse Approach}

\author{\IEEEauthorblockN{Shijian Gao, Xiang Cheng, \IEEEmembership{Senior~Member,~IEEE}, Liuqing Yang, \IEEEmembership{Fellow,~IEEE}}
\thanks{Part of this paper has been presented in IEEE International Conference on Communications, Shanghai, China,
 May 20-24, 2019\cite{0}.}}
\maketitle
\begin{abstract}
In mmWave massive multiple-input multiple-output (mMIMO) systems, hybrid digital/analog beamforming
has been recognized as an economic means to overcome the severe mmWave propagation loss.
To facilitate beamforming for mmWace mMIMO, there is a great urgency to acquire
accurate channel state information. To this end,
a novel doubly-sparse approach is proposed to estimate doubly-selective mmWave channels under hybrid mMIMO.
Via the judiciously designed training pattern, the well-known beamspace sparsity along with the
 under-investigated delay-domain sparsity that mmWave channels exhibit can be jointly exploited to assist channel estimation.
Thanks to our careful two-stage (random-probing and steering-probing) design, the proposed channel estimator
  possesses strong robustness against the double (frequency and time) selectivity whilst enjoying the benefits brought by the exploitation of double sparsity.
Compared with existing alternatives, our proposed channel estimator not only proves to be more general, but
 also largely reduces the training overhead, storage demand as well as computational complexity.
\end{abstract}

\section{Introduction}
Thanks to the abundant frequency resources at millimeter-wave (mmWave) band, mmWave communications have been recognized as one of key technologies for the 5G \& beyond wireless systems \cite{c1}\hspace{-0.01cm}-\hspace{-0.01cm}\cite{c2}. However, a major concern impeding the wide deployment of mmWave systems comes from the severe propagation loss\cite{c3}\hspace{-0.01cm}-\hspace{-0.01cm}\cite{c4}.
Fortunately, the signal attenuation can be compensated for by the large array gains, as the much shorter mmWave wavelength allows massive antenna arrays to be employed at the mmWave transceivers
\cite{c6}\hspace{-0.01cm}-\hspace{-0.01cm}\cite{c7a}.

To facilitate beamforming, the top priority is {to} acquire an accurate channel state information (CSI)\cite{c7b}. However, compared to the centimeter-wave (cmWave) MIMO systems,  channel estimation for mmWave mMIMO
 faces unprecedented challenges.
First, mmWave mMIMO typically adopts a hybrid structure for the power consumption and hardware cost concerns\cite{c4},\cite{c7c}, hence the large-scale channel matrix has to be estimated via a very limited number of RF chains. Since the latter essentially determines the number of
effective training symbols that can be transmitted simultaneously, {it can take significant amount of time to transmit} sufficient training symbols for mMIMO.
When it comes to the mobile scenarios, the problem becomes even more challenging, because {the channel turns out to be
time-varying in the presence of Doppler shifts.}

As mmWave channels exhibit limited scattering, a unique sparsity holds in beamspace under mMIMO. Thanks to this sparsity,
it may not be necessary to estimate the channel matrix element by element.
 Instead, one can resort to the compressed sensing (CS) theory to reduce the training overhead while ensuring a high accuracy.
 {Following this idea,} in \cite{c8}, a hierarchical beam training coupled with orthogonal matching pursuit (OMP) \cite{c8a} is devised to estimate static narrowband mmWave channels.
 In \cite{c8b}, block-OMP (BOMP) is applied to estimate narrowband \& time-varying mmWave channels.
 The static wideband channel estimation in the line-of-sight (LoS) scenarios is considered in\cite{c9},
 and the relevant work has been further extended to the non-LoS (NLoS) scenarios like \cite{c10} and \cite{c11},
 where OMP is applied either in the time-domain or the frequency-domain to assist channel estimation.

 Due to the wideband nature of mmWave, the narrowband channel model suffers from severe limitations, motivating us to focus on the wideband channel model.
 Generally speaking, {existing wideband channel estimation works} can be divided into two main categories:
   time-domain estimation vs. frequency-domain.
  The former is to estimate all channel taps jointly, while the latter is to estimate individual subcarriers independently.
  By exploiting the sparsity in beamspace, both schemes achieve similar performance in the sense of the normalized mean square error (NMSE),
  {with a largely reduced training overhead compared to the least-squares (LS) estimator.}
  However, to effectively exploit the sparsity so that OMP could be applied,
  either the demanding storage requirement or the heavy computational burden is inevitable. On top of that,
  these works have not taken into account of the {Doppler effects}, rendering their feasibility in mobility scenarios questionable.
To address these issues, there is an urgent need for a more generalized and more efficient channel estimation approach.

 To achieve significant reduction in training overhead and computational complexity,
  we resort to the under-exploited delay-domain sparsity in combination with the well-known beamspace sparsity.
  Motivated by the promising merits, a novel channel
 estimator is proposed by exploiting the double sparsity.
  As a matter of fact, the idea of using either the delay-domain sparsity or the double sparsity can be also found in some works, such as \cite{c11a}, \cite{c11b}, {and \cite{c11c}.}
However, these works are not specifically designed for hybrid mMIMO, and their studied channels have not taken time selectivity into account. In fact, once the time selectivity is involved, how to exploit either the delay-domain {sparsity or the} beamspace sparsity becomes a thorny problem. Furthermore, the introduction of the hybrid structure makes channel estimation a totally different topic as before, because the design flexibility is severely restricted by the hardware {constraints}.

Aiming at seeking a high-performance and easy-to-implement channel estimator, a doubly-sparse approach is innovatively proposed,
which can not only exploit the double sparsity to assist channel estimation, but can also provide a strong robustness to overcome the double selectivity.
The doubly-sparse approach comprises the following {steps}:
 \begin{itemize}
 \item To deal with the sparsity in delay domain, a special training pattern is judiciously designed to successfully separate each channel tap
 {regardless of Doppler effects}.
 Based on the energy detector, only a small proportion of channel taps will be identified effective and awaits a further processing.
\item To deal with {beamspace sparsity, an enhanced OMP} algorithm termed as A-BOMP is proposed to
{recover} the angle support. {A-BOMP can adaptively adjust basis matching \& residue update with properly determined iterations.
 With this, a high accuracy can be guaranteed even under strong Doppler effects}.
\item To jointly estimate the path gain and Doppler, a beamforming polling strategy is proposed based on the estimated angle supports. Via a few high-quality received samples after beamforming, both the path gain and Doppler can be reliably estimated with low training overhead.
 \end{itemize}

\noindent Compared with existing work, the doubly-sparse approach can remarkably improve the estimation accuracy,
 and largely reduce the training overhead, storage demand as well as  computational complexity.
 As many implementing issues are also considered in the specific design, the proposed channel estimator has a great potential to be applied in practice.

The rest of this paper is organized as follows: Section \uppercase\expandafter{\romannumeral2} describes the system and channel models.
 Section \uppercase\expandafter{\romannumeral3} and Section \uppercase\expandafter{\romannumeral4} introduce how to exploit the
  beamspace and delay-domain sparsity, respectively.
  Section \uppercase\expandafter{\romannumeral4} explains the estimation of the path gain and Doppler.
 Extensive numerical results and discussions are presented in Section \uppercase\expandafter{\romannumeral5},
  followed by conclusions in Section \uppercase\expandafter{\romannumeral6}.\\

\begin{figure}[t]
\centering
    \centering
    \includegraphics[width=3.5in]{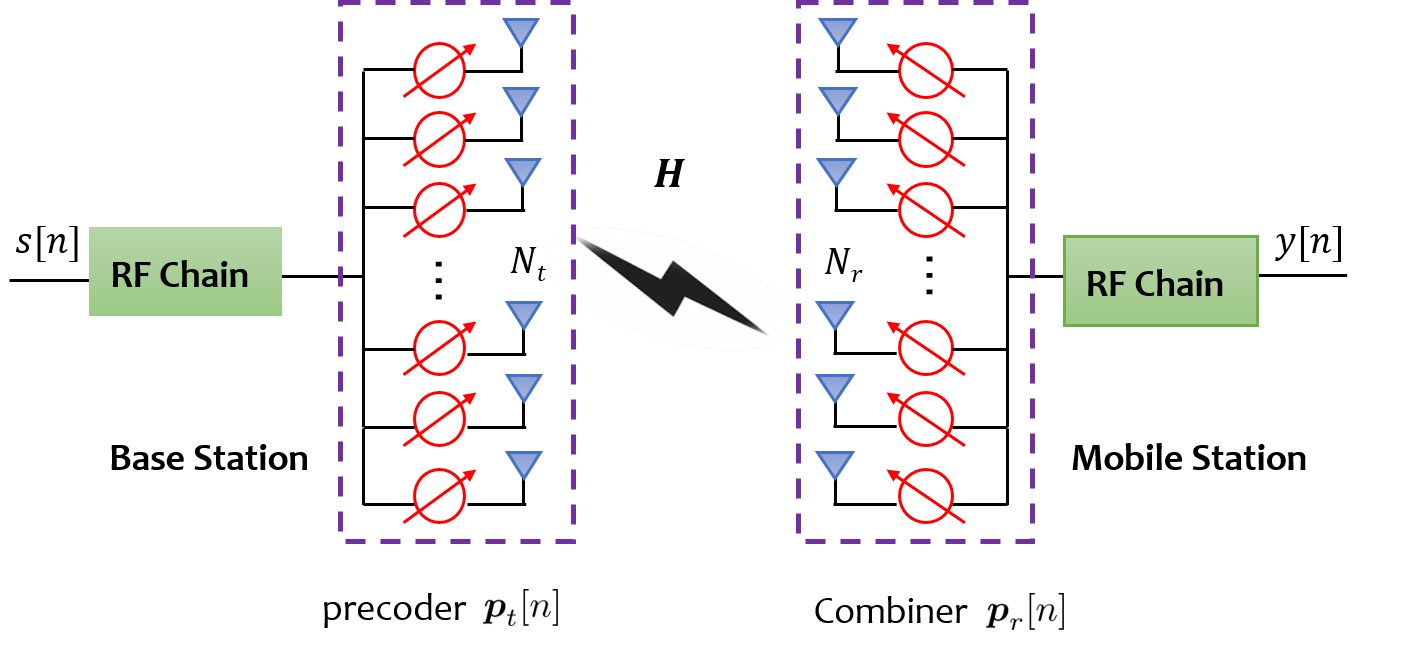}
    \caption{\small The system model of mmWave mMIMO transceivers with one RF chain}\label{Fig.1}
\end{figure}
\noindent{\em Notation:} In the remainder of the paper, $a$, $\bba$ and $\bbA$ represent a scalar, a vector and a matrix, respectively.
$|\mathcal{A}|$ is the cardinality of the support $\mathcal{A}$.
$\bbA[m,n]$, $\bbA[m,:]$ and $\bbA[:,m]$ are denoted as the $(m,n)$ entry, the $m$-th row, and the $m$-th column of $\bbA$, respectively.
$\bbA^{'}$, $\bbA^*$ and $\bbA^{\dagger}$ denote the transpose, the hermitian transpose and the pseudo-inverse of $\bbA$, respectively.
 ${\left\lfloor  \cdot  \right\rfloor }$ and $\lceil \cdot \rceil$ represent the floor and ceiling operation, respectively.
{\textbf{cal}(.) represents the cardinality. \textbf{diag}(.) and \textbf{vec}(.) represent the operations of
 diagonalization and vectorization, respectively.} $\mathbb{E}$ stands for expectation.  ${\mathcal {CN}(0,\sigma^2)}$ represents the distribution of a circularly symmetric complex Gaussian random value with variance $\sigma^2$.

\section{System and Channel Description}
\subsection{System Model}
A mmWave mMIMO system is considered,
where $N_t$ and $N_r$ antennas are employed at the transmitter (Tx) and receiver (Rx), respectively. Since the proposed channel estimation approach does not rely on
the channel reciprocity, we simply assume channel estimation is implemented at Rx.
For the power consumption and hardware cost concerns, mmWave mMIMO typically adopts a hybrid structure,
in which the number of RF chains at the transceivers is much smaller than that of the antennas.
 Similar to \cite{c4}\cite{c12}, a fully-connected hybrid structure is studied here, where
 the RF chains and antennas are connected via a digitally controlled analog phase shifter (APS) network.
Suppose each APS component has a resolution of $b$ bits, then all adjustable angles are contained in
\begin{align}
\mathcal{B}=\left\{0,{2\pi}/{2^b},\cdots,{2\pi(2^b-1)}/{2^b}\right\}
\end{align}
with cardinality $\mid\hspace{-0.1cm}\mathcal{B}\hspace{-0.1cm}\mid=2^b$. {Accordingly, the angular quantization function is expressed as
\begin{align}
\mathcal{Q}(x)=\mathcal{B}(i^{*}), i^*=\mathop{\arg\min\limits_{i}}\mod\big(x-\mathcal{B}(i),2\pi\big).
\end{align}}
As in \cite{c12_1}, let the transceivers each employ a single RF chain as shown in Fig.~1.
Note that, since
we focus on channel estimation in this paper, this setup is without loss of generality and can be readily generalized to cope with arbitrary number of RF chains
at the transceivers.

\subsection{Geometric Channel}
In this paper, we adopt the modified Sen-Matolak channel model\cite{c12a}-\hspace{-0.02cm}\cite{c12b}, which is an extension of the narrowband geometric model by taking the path delay and the Doppler effect into account.
Denote the maximum number of delay taps as $N_c$. {At time instant $n$, the sampled version of the tap-$d$ channel $(0\leq d< N_c)$ is given by}
\begin{align}\label{a1}
\bbH_d(n)\hspace{-0.1cm}=\hspace{-0.1cm}\sum_{p=1}^{P}\hspace{-0.05cm}\sqrt{\frac{N_tN_r}{P}}\alpha_ph(dT_s\hspace{-0.05cm}-\hspace{-0.05cm}\tau_p)\bba_r(\theta_p)\bba^*_t(\phi_p)e^{j\omega_pn}
\end{align}
where $\alpha_{p}\sim \mathcal{CN}(0,1)$ is the complex gain of the $p$-th path; $h(\cdot)$ is the pulse shaping filter response;
 $\tau_p$ is the propagation delay of the $p$-th path that obeys a uniform distribution on $[0,(N_c-1)T_s)$;
  $\theta_p$ and $\phi_p$ represent the angle of arrival (AoA) and angle
of departure (AoD), respectively, both of which being modeled as uniformly distributed variables on $[0,2\pi)$.
{Define the system carrier frequency to be $f_c$, the velocity of light to be $c_v$, and the maximum relative velocity to be $v_m$.
 Then the normalized Doppler shift is $\omega_p=2\pi{f_cv_mT_s\sin(\theta_p)}/{c_v}$}.
For notational simplicity, an array response generating function is defined as
\begin{align}
\bbf_N(y)=\frac{1}{\sqrt{N}}\big[1,e^{j2\pi y},\cdots,e^{j2\pi(N-1)y}\big]^{'}.
\end{align}
With half-wavelength uniform linear arrays (ULAs) employed at the transceivers, we have { $\bba_t(\phi)=\bbf_{N_t}\big(\sin(\phi)/2\big)$ and $\bba_r(\theta)=\bbf_{N_r}\big(\sin(\theta)/2\big)$.}

\subsection{Beamspace Representation}
To simplify expression, one can rewrite the geometric model into the following compact form{\cite{c8b}}
\begin{align}\label{cf}
\bbH_d(n)=\bbA_R \mathbf{diag}\big(\bbg_d(n)\big)\bbA^*_T
\end{align}
where $\bbA_T=\big[\bba_t(\phi_1),\bba_t(\phi_2),\cdots,\bba_t(\phi_P)\big]\in \mathcal{C}^{N_t\times P}$ and $\bbA_R=\big[\bba_r(\theta_1),\bba_r(\theta_2),\cdots,\bba_r(\theta_P)\big]\in \mathcal{C}^{N_r\times P}$
are the transmit and receive steering matrices, respectively, {which remain unchanged during the channel estimation stage. The time-varying effects are incorporated in $\bbg_d(n)$ given by
\begin{align}
\sqrt{\frac{N_tN_r}{P}}\big[\alpha_1h(dT_s-\tau_1)e^{j\omega_1n},\cdots,\alpha_Ph(dT_s-\tau_P)e^{j\omega_Pn}\big]^{'}
\end{align}
which contains the path gains  at time instant $n$}.

In Eq. (\ref{cf}), $\bbA_T$, $\bbA_R$ as well as $\bbg_d(n)$  are all associated with the physical channel taps, which
are not always resolvable due to the finite resolution of the receiver in time and space, and thus cannot be directly estimated.
To seek an equally general but more practical representation, we first {construct the Tx-end and Rx-end angular dictionary matrices as in \cite{c8}}
\begin{subequations}
\begin{align}
&\bbD_t=\big[\bbf_{N_t}(0),\bbf_{N_t}({1}/{G_t}),\cdots,\bbf_{N_t}\big({(G_t\hspace{-0.05cm}-\hspace{-0.05cm}1)}/{G_t}\big)\big] \label{a6_1} \\
&\bbD_r=\big[\bbf_{N_r}(0),\bbf_{N_r}({1}/{G_r}),\cdots,\bbf_{N_r}\big({(G_r\hspace{-0.05cm}-\hspace{-0.05cm}1)}/{G_r}\big)\big]\label{a6_2}
\end{align}
\end{subequations}
where $G_t$ and $G_r$ represent the size of corresponding dictionaries.
Taking $\bbD_t$ as an example, it contains the steering vectors ranging from $[0,2\pi]$ with resolution $2\pi/G_t$.
{As $G_t$ approaches infinity, the resolution becomes zero, thus leading to a continuous dictionary}. For practical implementation,
 most work show that setting $G_t$ as $2\sim 4$ times the array size can provide sufficient resolution for separating the AoAs/AoDs of the propagation paths.
Based on the dictionary matrices, the channel representation in Eq. (\ref{cf}) can be re-expressed as
\begin{align}\label{br}
\bbH_d(n)=\bbA_R \mathbf{diag}\big(\bbg_d(n)\big)\bbA^*_T=\bbD_r\overline{\bbH}_d(n)\bbD^*_t.
\end{align}

 Under the mMIMO setup, $P$ propagation paths result in $P$ dominant non-zero entries in $\overline{\bbH}_d$.
  As $\bbD_r$ and $\bbD_t$ are irrelevant to $\bbH_d$, $\overline{\bbH}_d$ essentially
  gathers the entire channel information that was originally contained by $\bbA_T$, $\bbA_R$ and $\bbg_d(n)$.
  Specifically, by omitting the time instant and assuming on-grid AoA/AoD pairs,
  $\forall p\in [1,P]$, $n_p=\frac{\phi_p}{2\pi/G_t}$, $m_p=\frac{\theta_p}{2\pi/G_t}$, $\overline{\bbH}_d(m_p,n_p)=\bbg_d[p]$.
 Therefore, $\overline{\bbH}_d$ is the channel representation in beamspace.
 Due to the limited scattering in mmWave propagation, $P\ll N_tN_r< G_rG_t$  \cite{c12}, {thus $\overline{\bbH}_d$
 can be recognized as a sparse matrix with sparsity $P$.}\footnote{In practice, the off-grid leakage arising from the finite resolution of beamspace may
 lead to extra non-zero entries in $\overline{\bbH}_d$.
Since the leakage is typically very weak under mMIMO, the ensemble of dominant entries in $\overline{\bbH}_d$ is still similar to $\bbg_d$.
Regardless of whether the non-zero entries of $\overline{\bbH}_d$ corresponds to a single channel path $p$ or to some leakage terms, these entries
are the resoluble ones that can be estimated. Even in the rare cases of strong leakages, the sparsity will not be affected since $P\ll G_tG_r$.}

{Revisiting Eq.~(\ref{a1}), the prior information available at both ends
are $N_t$, $N_r$, $N_c$, the steering pattern of $\bba_{t/r}$, while the remaining parameters are unknown to the transceivers, and thus
have to be recovered via channel
estimation. In the following, we will heavily rely on the beamspace representation to recover the beam direction (AoA \& AoD), the beam amplitude, as well
as the associated Doppler shift.}

\subsection{{Input-output} relationship}
Let $s(n)$ be the training symbol at instant-$n$. At the Tx, $s(n)$ is first processed at the APS network,
and the transmitted signal is
$\bbx(n)=\bbp_t(n)s(n)\in \mathcal{C}^{N_t\times 1}$. Since each APS component can only adjust the phase,
the probing vector $\bbp_t(n)$ bears the form as
\begin{align}\label{pbv}
\bbp_t(n)=\sqrt{{1}/{N_t}}\big[e^{j\alpha_1(n)},e^{j\alpha_2(n)},\cdots,e^{j\alpha_{N_t}(n)}\big]^{'}
\end{align}
with $\alpha_i(n)\in \mathcal{B}$, $\forall i\in[1,N_t]$.

After channel propagation, the received signal is
\begin{align}
\bbr(n)=\sum_{d=0}^{N_c-1}\bbH_d(n)\bbx(n-d)+\bm{\eta}(n)
\end{align}
which is the convolution of multiple time-varying channel taps. $\bm{\eta}(n)\sim\mathcal{CN}(\mathbf{0},\sigma^2\bbI_{N_r})$ is the receiver noise vector.
$\bbr(n)$ then goes through the Rx-end APS network, whose function is described by an $N_r\times 1$ probing vector $\bbp_R(n)$,
 so the received sample after APS becomes
\begin{align}\label{IO}
y(n)=\hspace{-0.1cm}\sum_{d=0}^{N_c-1}\bbp^*_r(n)\bbH_d(n)\bbp_t(n\hspace{-0.05cm}-\hspace{-0.05cm}d)s(n\hspace{-0.05cm}-\hspace{-0.05cm}d)+\xi(n).
\end{align}
where  $\xi(n)=\bbp^{*}_r(n)\bm{\eta}(n)\sim\mathcal{CN}(0,\sigma^2)$ remains white.
{Let $\overline{\bbp}_t(n)=\bbD_t\bbp_t(n)$ and $\overline{\bbp}_r(n)=\bbD_r\bbp_r(n)$. Based on the beamspace representation in Eq. (\ref{br}),} we have
\begin{align}
y(n)=\hspace{-0.1cm}\sum_{d=0}^{N_c-1}\overline{\bbp}^*_r(n)\overline{\bbH}_d(n)\overline{\bbp}_t(n\hspace{-0.05cm}-\hspace{-0.05cm}d)s(n\hspace{-0.05cm}-\hspace{-0.05cm}d)+\xi(n).
\end{align}

Without loss of generality, we consider the general I-O relationship for the first frame only
 unless otherwise specified. {Here the length-$N_f$ training frame is simply denoted as $[s(0),s(1),\cdots,s(N_f-1)]$, and its specific form will be explained later.}
By concatenating all received samples, {we write the I-O relationship in matrix form shown in Eq.~(\ref{cio}) at the top of next page,}
\begin{figure*}
\begin{align}\label{cio}
\bby&=[y(0),y(1),\cdots,y(N_f-1)]^{'}\nonumber\\
&={\overline{\bbP}^*_r}\left[
    \begin{array}{ccccc}
      \overline{\bbH}_0(0) & \mathbf{0} & \mathbf{0} & \cdots & \mathbf{0} \\
      \vdots & \overline{\bbH}_0(1) & \mathbf{0} & \cdots & \mathbf{0} \\
      \overline{\bbH}_{N_c-1}(N_c-1) & \cdots & \ddots & \cdots & \vdots \\
      \vdots & \ddots & \cdots & \ddots & \mathbf{0} \\
      \mathbf{0} & \vdots &  \overline{\bbH}_{N_c-1}(N_f-1) & \cdots & \overline{\bbH}_0(N_f-1) \\
    \end{array}
  \right]\overline{\bbP}_t\left[
                                                \begin{array}{c}
                                                  s(0) \\
                                                  \vdots \\
                                                  s(N_c\hspace{-0.1cm}-\hspace{-0.1cm}1) \\
                                                  \vdots \\
                                                  s(N_f\hspace{-0.1cm}-\hspace{-0.1cm}1) \\
                                                \end{array}
                                              \right]+\bm{\xi}.
\end{align}
\end{figure*}
with
$\overline{\bbP}^*_r=\hspace{-0.15cm}
\left[\hspace{-0.1cm}
  \begin{array}{cccc}
    \overline{\bbp}^*_r(0) & \mathbf{0} & \cdots & \mathbf{0} \\
    \mathbf{0} & \overline{\bbp}^*_r(1) & \cdots& \mathbf{0} \\
     \vdots& \vdots & \ddots & \mathbf{0} \\
    \mathbf{0}& \mathbf{0} & \cdots & \overline{\bbp}^*_r(N_f\hspace{-0.05cm}-\hspace{-0.05cm}1) \\
  \end{array}
\hspace{-0.2cm}\right]$
and $\overline{\bbP}_t=\hspace{-0.15cm}
\left[\hspace{-0.1cm}
  \begin{array}{cccc}
    \overline{\bbp}_t(0) & \mathbf{0} & \cdots & \mathbf{0} \\
    \mathbf{0} & \overline{\bbp}_t(1) & \cdots& \mathbf{0} \\
     \vdots& \vdots & \ddots & \mathbf{0} \\
    \mathbf{0}& \mathbf{0} & \cdots & \overline{\bbp}_t(N_f\hspace{-0.05cm}-\hspace{-0.05cm}1) \\
  \end{array}
\hspace{-0.2cm}\right]$.

\section{Exploit delay-domain sparsity}
As described in Section \uppercase\expandafter{\romannumeral2}, mmWave channels exhibit sparsity in beamspace.
Apart from this well-known sparsity,
this section will further show that mmWave channels exhibit sparsity in the delay domain as well.
We first analyze why existing approaches fail to exploit the delay-domain sparsity,
and then explain how one can effectively benefit this largely overlooked sparsity.

\begin{figure}[htb]
\centering
    \centering
    \includegraphics[width=3.3in,height=2.4in]{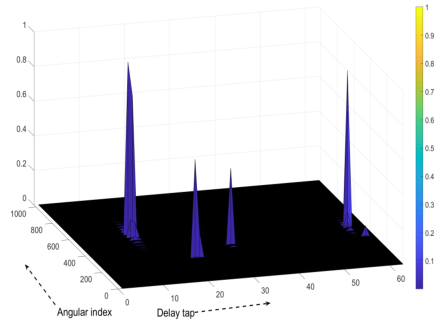}
    \caption{\small The delay-beamspace colormap of a randomly generated mmWave channel with {$N_t=N_r=32$, $N_c=64$ and $P=5$}.}\label{Fig.5}
\end{figure}

\subsection{Sparsity in delay domain}
To eliminate inter-frame interferences (IFIs) in block transmission, a commonly adopted approach amounts to zero-padding (ZP)
a guard interval with length at least $(N_c-1)$ to each frame.
 For example, the data-frame length is $512$ in IEEE 802.11ad, while the prefix length can be up to $128$ \cite{c13}. However,
a long delay spread with large $N_c$ due to the high symbol rate does not mean a rich multi-path environments.
 In fact, mmWave channels have very few dominant paths\footnote{typically $8\sim12$ even in ``rich'' scattering environments, and is much less in other environments \cite{c13c}. Similar evidences can also be found in mainstream standards (see, e.g.,\cite{c13} and \cite{c16}).}.
 Hence, a majority of the channel taps are actually too weak to be considered, rendering  sparsity in the delay domain.
  To gain some intuitive insight, we plot the colormap of a randomly generated channel in Fig. 2, where the double sparsity
  in both the beamspace and delay domain can be clearly observed.
\subsection{Conventional Training Pattern}
The currently adopted training pattern\cite{c11} is given by
{
\begin{align}
&[s(0),s(1),\cdots,s(N_f-1)]\nonumber \\
=&\big[s_0,s_1,s_2,\cdots,s_{N-1},\underbrace{0,\cdots,0}_{N_c}\big].
\end{align}}
Specifically, each frame contains $N_f=N+N_c$ symbols,
 where $N$ and $N_c$ are the length of the data sequence and ZP, respectively.
Clearly, the I-O relationship of this pattern still follows the general one in Eq. (\ref{cio}),
 but some specifics need to be clarified.

In wideband mmWave systems, symbols are pumped out at a very high rate, thus leaving
insufficient buffer time for the APS network reconfiguration except for the ZP interval\cite{c10}.
As a result, the probing vectors remain unchanged over the entire frame, that is{
\begin{align}
\bbp_{t/r}(n)=\bbp_{t/r}(0),~ \forall n\in[0,N_f).
\end{align}
Accordingly, $\overline{\bbP}^*_r=\bbI_{N_f}\otimes\overline{\bbp}^*_r(0)$ and $\overline{\bbP}_t=\bbI_{N_f}\otimes\overline{\bbp}_t(0)$}.

Although the introduction of ZP ensures IFI-free, $N_c$ channel taps remain unresolvable after convoluting with the training sequence.
In consequence, channel estimation requires joint processing across all taps, leading to high storage demand and heavy computational burden.
More importantly, {exploiting the delay-domain sparsity becomes an intractable task.}

\subsection{Proposed training pattern}
To avoid these limitations, there is a great urgency to devise a new training pattern, by which the delay-domain sparsity could be exploited to facilitate channel estimation, {and the pattern itself must be friendly to implementation.} To this end, a new training pattern is designed as follows
{
\begin{align}\label{a2}
&[s(0),s(1),\cdots,s(N_f-1)]\nonumber \\
=&\big[\underbrace{s_0,\overbrace{0,\cdots,0}^{N_c-1}}_{
(1)}\bigg|\underbrace{s_1,\overbrace{0,\cdots,0}^{N_c-1}}_{
(2)}\bigg|\cdots\cdots\bigg|\underbrace{s_{L-1},\overbrace{0,\cdots,0}^{N_c-1}}_{
(L)}\big].
\end{align}
}
As can be seen,  each frame is further divided into $L={N_f}/{N_c}$ subframes\footnote{Without loss of generality, $L$ is assumed to be an integer here.
If ${N_f}/{N_c}$ is not an integer, one can simply use $L=\lfloor{N_f}/{N_c}\rfloor$.}.
Owing to the ZP in each subframe,
sufficient buffer time is left to reconfigure the APS network after each non-zero training symbol. In other words,
  the probing vectors can be updated $L$ times per frame, i.e.,
  \begin{align}\label{GPV}
 \bbp_{t/r}(n)=\bbp_{t/r}\big(N_c\lfloor{n}/{N_c}\rfloor\big), \forall n\in[0,N_f).
 \end{align}
 As an interesting fact,
 when it comes to the conventional frequency-selective MIMO channel estimation,
  this training pattern has been proved optimal in the sense of both the mean squared error (MSE) and  end-to-end mutual information\cite{c13b}.

Before taking a closer look at the I-O relationship using the new pattern, we first make the following definition.

{
\noindent \textit{\textbf{Random-probing vector}: At the random-probing stage, the probing vectors are generated by randomly adjusting the angle of each APS component from $\mathcal{B}$. The resultant vector is termed as the random-probing vector and denoted as
\begin{align}
{\bbp}^R_{t/r}(l)={\bbp}_{t/r}(lN_c+n_c), (l<L, n_c<N_c).
\end{align}}}
{Applying random probing is simply because no prior CSI is available at this stage.
Note that, the above definition implies that ${\bbp}^R_{t/r}$ possesses both the randomness and subframe-updatability property.
Applying a similar notational change to $\overline{{\bbp}}_{t/r}$, $\overline{\bbP}^*_r$ becomes Eq.~(\ref{newpr}),
 \begin{figure*}
 \begin{align}\label{newpr}
 \overline{\bbP}^*_r=\left[
  \begin{array}{cccc}
    \bbI_{N_c}\otimes\big(\overline{{\bbp}}^R_{r}(0)\big)^* & \mathbf{0} & \cdots & \mathbf{0} \\
    \mathbf{0} &  \bbI_{N_c}\otimes\big(\overline{{\bbp}}^R_{r}(1)\big)^* & \cdots& \mathbf{0} \\
     \vdots& \vdots & \ddots & \mathbf{0} \\
    \mathbf{0}& \mathbf{0} & \cdots &  \bbI_{N_c}\otimes\big(\overline{{\bbp}}^R_{r}(L-1)\big)^* \\
  \end{array}
\right]
\end{align}
\end{figure*}
and $\overline{\bbP}_t$ is obtained likewise. Substituting the new $\overline{\bbP}^*_r$ and $\overline{\bbP}_t$,
together with the training frame into Eq. (\ref{cio}), the received signal becomes}
\begin{align}\label{nyn}
&y(lN_c+n_c)\nonumber\\
=&\big(\overline{\bbp}^R_r(l)\big)^*\overline{\bbH}_{n_c}(lN_c+n_c)\overline{\bbp}^R_t(l)s_{l}+\xi(lN_c+n_c).
\end{align}
Clearly, the received samples are now associated with a single channel tap.
Hence, our proposed pattern facilitates separating channel taps and thus rendering it possible to exploit the delay-domain sparsity easily.
Since the success of tap separation does not rely on the non-zero training symbol $s_l$ in Eq.~(\ref{a2}), $s_l$ is set as 1 in the rest of paper without loss of generality.

\subsection{Identification of effective taps}
To determine the existence of tap-$d$ channel, we gather all $\overline{\bbH}_d$-related samples, i.e., $y(lN_c+d)$, $\forall l\in[0,L-1]$.
 If at least one dominant path exists in the tap-$d$ channel, $y(lN_c+d)$ includes both the signal and noise parts.
Otherwise, $y(lN_c+d)$ contains noise only. Hence, detecting the existence of the tap-$d$ channel is
 a binary hypothesis testing problem that can be dealt with via energy detector. We first average the power of all samples associated with $\overline{\bbH}_d$ \footnote{{In this paper, we assume one frame is sufficiently long, so the channel estimator is explained based on one frame. In practice, one can simply replace $L$ with the actual number of subframes used at the random-probing
stage.}}, and get the test statistics (TS) and normalized TS (nTS) as
\begin{subequations}
\begin{align}\label{TS}
Y_d&=\frac{1}{L}\sum_{l=0}^{L-1}\big|y(lN_c+d)\big|^2\\
\overline{Y}_d&=\frac{Y_d-\sigma^2}{\max\limits_{0\leq m<N_c}(Y_m-\sigma^2,0)}.
\end{align}
\end{subequations}
{When applying CS, random probing is necessary in estimating both the time-invariant and time-varying channels.}
While for the latter, another important function of random probing is to {remain robust against Doppler.}

\textbf{Proposition 1}~[Validity of test statistics with Doppler]: \textit{With sufficient random probings, the test statistics $Y_d$
is approximately irrelevant to the channel's time variation.}

{\it proof}: Let $n=lN_c+d$ and {$g_{d,p}(n)$} be the $p$-th element of {$\bbg_{d}(n)$}, then
\begin{align}
y(n)&=\sum_{d=0}^{N_c-1}\overline{\bbp}^*_r(n)\overline{\bbH}_d(n)\overline{\bbp}_t(n\hspace{-0.05cm}-\hspace{-0.05cm}d)+\xi(n)\nonumber\\
&=\hspace{-0.1cm}\sum\limits_{p=1}^P\hspace{-0.1cm}\big(\bbp^R_r(l)\big)^*\bba_r(\theta_p)\bba^*_t(\phi_p)\bbp^R_t(l){g_{d,p}(n)}\hspace{-0.05cm}+\hspace{-0.05cm}\xi(n).
\end{align}
 Denote $\rho_p(l)=(\bbp^R_R(l))^*\bba_r(\theta_p)\bba^H_t(\phi_p)\bbp^R_T(l)$.  By using {$g_{d,p}(n)=g_{d,p}(0)e^{j\omega_{p}n}$},
we have
\begin{align}\label{dets}
&|y(n)|^2\hspace{-0.1cm}=\hspace{-0.1cm}\sum\limits_{p=1}^P\hspace{-0.05cm}|\rho_p(n)g_{d,p}(0)|^2\hspace{-0.05cm}+\hspace{-0.05cm}2\mathcal{R}\bigg\{\hspace{-0.1cm}\sum\limits_{p=1}^P\rho_{p}(l)g_{d,p}(0)\xi(n)e^{j\omega_{p_1}n}\hspace{-0.05cm}\bigg\}\nonumber\\
&+2\mathcal{R}\bigg\{\sum_{p_1}\sum_{p_2}\rho_{p_1}(l)g_{d,p_1}(0)\rho^*_{p_2}(l)g^*_{d,p_2}(0)e^{j(\omega_{p_1}-\omega_{p_2})n}\bigg\}.
\end{align}
Since $\bbp^R_t$ and $\bbp^R_r$ are random probing vectors with zero mean, it can be readily verified $\mathbb{E}\{{\rho_{p}(l)}\}=0$, $\forall p\in[1,P]$.
By averaging sufficient $|y(n)|^2$ terms, the last two terms in Eq. (\ref{dets}) approach zero, thus the TS becomes irrelevant to $\omega_{p}$.

Proposition 1 guarantees {the exploitation of the delay-domain sparsity regardless of Doppler effects}. With the energy detector, the effective taps can be roughly selected as
\begin{align}
\mathcal{P}_1=\left\{d~\bigg|~\overline{Y}_d\geq \mu \right\}\bigcap\left\{d~\bigg|~Y_d>\sigma^2\right\}
\end{align}
where $\mu$ is the threshold\footnote{{To balance the accuracy and complexity,
it is suggested to set $\mu$ several dBs smaller than the anticipated normalized MSE defined as $\varepsilon={\sum_{d=0}^{N_c-1}\parallel\bbH_d-\widehat{\bbH}_d\parallel_F}/{\sum_{d=0}^{N_c-1}\parallel\bbH_d\parallel_F}$, where $\widehat{\bbH}_d$ is the estimated channel.
Evidently, the design of energy detector is heuristic due to the difficulty of getting the distribution of $y(n)$.}}.
To avoid extreme cases where $\mathbf{cal}(\mathcal{P}_1)$ is either 0 or unreasonably large,
{a tuning procedure is added, and the ultimately determined taps are given by}
\begin{align}\label{cr1}
\mathcal{P}\hspace{-0.05cm}=\hspace{-0.05cm}
\begin{cases}
~~~~~~\mathcal{P}_1,~~~0<\mathbf{cal}(\mathcal{P}_1)\leq A\\
\big\{d|\overline{Y}_d\geq \overline{\lambda}_A\big\},~\mathbf{cal}(\mathcal{P}_1)> A\\
\big\{d|Y_d\geq \lambda_A\big\},~~\mathbf{cal}(\mathcal{P}_1)=0.
\end{cases}
\end{align}
with $\lambda_A$ and $\overline{\lambda}_A$ representing the $A$-th largest TS and nTS, respectively.

Up till now, we have accomplished {the first part of the random-probing stage}. Summarizing, the main steps can be described as follows:
\begin{itemize}
\item Transmit judiciously designed training frames with random APS probing.
\item Calculate the TS/nTS for each channel tap based on the corresponding received samples.
\item Determine the non-negligible channel taps based on the energy detector $\mathcal{P}$.
\end{itemize}

\section{Exploiting the Beamspace Sparsity}
As outlined in Section \uppercase\expandafter{\romannumeral2}, the beamspace channel exhibits sparsity under mMIMO settings. Therefore, instead of
estimating the {original geometric channel matrix $\bbH_d$ with dimension $N_tN_r$}, we estimate the sparse beamspace channel $\overline{\bbH}_d$. Since the time variation imposes a great difficulty in recovering the exact values of non-zero entries from $\overline{\bbH}_d$, we focus on locating the non-zero entries (essentially the angle support)
{first in this section, and leaving the estimation of exact values to the next section}.
\subsection{Sparse transformation}
After tap detection, $n_c$ out of $N_c$ taps are recognized effective, with their indices collected by
$\mathcal{D}=\{d_1,d_2,\cdots,d_{D}\}$.
{Using the samples already obtained at the random-probing stage}, we proceed to determine the angle support for those taps belonging to $\mathcal{D}$.
{It has to be stressed that this step does not require extra training frames.}

Due to the similarity, we take tap-$d_i$ ($d_i\in\mathcal{D}$) for example, and the subscript {of $d_i$} is omitted for brevity.
{To apply CS, let us first derive the sparse representation for received samples.}
Stacking all $\overline{\bbH}_d$-related samples from $\bby$ yields
\begin{align}
\bby_d=[y(d),y(N_c+d),\cdots,y((L-1)N_c+d)]^{'}.
\end{align}
Denoting $n_l=lN_c+d$ $\big(\forall l \in[0,L)\big)$ and using matrix equality $\mathrm{vec}(\bbA\bbB\bbC)\hspace{-0.1cm}=\hspace{-0.1cm}(\bbC^{'}\otimes\bbA)\mathrm{vec}(\bbB)$, $y(n_l)$ can be rewritten as
\begin{align}\label{a4}
y(n_l)=\underbrace{\bigg(\big(\overline{\bbp}^R_t(l)\big)^{'}\otimes\big(\overline{\bbp}^R_r(l)\big)^*\bigg)}_{\bm{\psi}(l)}\underbrace{\bigg(\mathbf{vec}\big(\overline{\bbH}_d(n_l)\big)\bigg)}_{\overline{\bbh}_d(n_l)}
+\xi(n_l).
\end{align}
Neglecting the noises temporarily for brevity, $\bby_d$ can be further expressed as
\begin{align}\label{bb}
\bby_d
    =\hspace{-0.15cm}\underbrace{\left[\hspace{-0.2cm}
        \begin{array}{cccc}
          \bm{\psi}(0) & \mathbf{0} & \cdots & \mathbf{0} \\
          \mathbf{0} & \bm{\psi}(1) & \cdots & \mathbf{0} \\
          \vdots & \vdots & \ddots & \vdots \\
          \mathbf{0} & \mathbf{0} & \cdots & \bm{\psi}(L-1) \\
        \end{array}
      \hspace{-0.15cm}\right]}_{\bm\Psi}\underbrace{\left[\hspace{-0.15cm}
         \begin{array}{c}
           \overline{\bbh}_d(n_0) \\
           \overline{\bbh}_d(n_1) \\
           \vdots \\
           \overline{\bbh}_d(n_{L-1}) \\
         \end{array}
      \hspace{-0.15cm}\right]}_{\overline{\bbh}_d}.
\end{align}

In the special case of time-invariant channels,  all $\overline{\bbh}_d(n_l)$'s are exactly the same\cite{c8b},
 thus giving rise to
\begin{align}\label{a3}
\bby_d={\left[
         \begin{array}{c}
           \bm{\psi}(0) \\
           \bm{\psi}(1) \\
           \vdots \\
           \bm{\psi}(L-1) \\
         \end{array}
       \right]}\overline{\bbh}_d(n_0).
\end{align}
Determining the angle support of $\overline{\bbH}_d$ is equivalent to locating  non-zero entries from the $G_tG_r$-dimensional vector $\overline{\bbh}_d$. Since $P\ll G_tG_t$, the ``localization'' can be effectively solved via OMP, through which $\mathcal{O}(P\log G_tG_r)$ instead of $\mathcal{O}(G_tG_r)$ samples are sufficient to guarantee a high accuracy.

However, {Eq. (\ref{a3}) is no longer valid in the presence of Doppler shifts, motivating us to restudy the more general Eq.~(\ref{bb})}.
 Because $\overline{\bbh}_d$ remains sparse for {$LP\ll LG_tG_r$},  a natural option would be OMP as well.
 Reminisce that
 the variations of AoAs/AoDs are negligible during the channel estimation, thus a common angle support is shared by all $\overline{\bbh}_d(n_l)$'s. However, OMP cannot exploit such a unique structure because it treats $\bbh_d(n_l)$ as a generic sparse vector.
 {Fortunately, by utilizing the unique property of $\overline{\bbh}_d(n_l)$, a more general block-sparse representation can be derived.} Specifically, constructing such a permutation matrix $\bbP$ satisfying
$\bbP[:,(i\hspace{-0.05cm}-\hspace{-0.05cm}1)G_tG_r\hspace{-0.05cm}+\hspace{-0.05cm}j\big]=\bbI_{G_tG_r}[:,(j\hspace{-0.05cm}-\hspace{-0.05cm}1)G_tG_r\hspace{-0.05cm}+\hspace{-0.05cm}i]$ \cite{c16a},
$\bby_d$ can be decomposed as
\begin{align}\label{a5}
\bby_d=\big({\bm\Psi}\bbP\big)\cdot\big(\bbP^{'}\overline{\bbh}_d\big).
\end{align}
The ``new'' sparse signal and sensing matrix then become
\begin{subequations}
\begin{align}
&{\widetilde{\bbh}_d}=\bbP^{'}\overline{\bbh}_d=\big[\widetilde{\bbh}^{'}_{d,1},\widetilde{\bbh}^{'}_{d,_2},\cdots,\widetilde{\bbh}^{'}_{d,G_tG_r}\big]^{'} \label{bvc}\\
&\bm{\widetilde{\Psi}}={\bm\Psi}\bbP=\big[\bm{\widetilde{\Psi}}_{1},\bm{\widetilde{\Psi}}_{2},\cdots,\bm{\widetilde{\Psi}}_{G_tG_r}\big]
\end{align}
\end{subequations}
where $\widetilde{\bbh}_{d,i}=\big[\overline{h}_{d,i}(n_0),\overline{h}_{d,i}(n_1),\cdots,\overline{h}_{d,i}(n_{L-1})\big]^{*}$ and
$\bm{\widetilde{\Psi}}_{i}=\mathbf{diag}\big[\psi_{i}(0),\psi_i(1),\cdots,\psi_i(L-1)\big]$, $\forall i\in[1,G_tG_r]$, with $\overline{h}_{d,i}(n_j)$ and $\psi_{i}(l)$ being the $i$-th entry of $\overline{\bbh}_d(n_j)$ and $\bm{\psi}(l)$, respectively.
Unlike the original $\overline{\bbh}_d$, the rearranged $\widetilde{\bbh}_d$ exhibits block sparsity\cite{c16b}. More importantly,
 the block sparsity of $\widetilde{\bbh}_d$ equals to the sparsity of $\overline{\bbh}_d(n_l)$.
Even with the block sparsity,
to accurately localize the non-zero support still encounters two major difficulties:
\begin{enumerate}[P.1]
\item How to properly set the number of iterations when applying CS algorithms.
\item How to avoid the potential degradation resulting from {the strong Doppler effects}.
\end{enumerate}
{To address these problems, we propose an algorithm termed as adaptive-block OMP (A-BOMP) that will be detailed next}.

\subsection{A-BOMP}
When recovering the sparse signal via CS, a proper number of iterations is equal to (or a slightly higher than) the signal sparsity. Unfortunately,
the actual sparsity of $\widetilde{\bbh}_d$ is unknown. To reduce the risks of estimation loss, most works adopt large iterations.
 However, once the iterations severely mismatch the signal sparsity, it may result in
increased computational complexity and potential over-fitting errors.
Albeit not knowing the sparsity either, we will show that, it is possible to set iterations properly after tap identification.

Since $D$ out of $N_c$ taps are regarded effective,
 the number of beams should be no greater than $D$, thus the signal sparsity is upper bounded by $D$.
Surprisingly, the upper bound could be set even smaller for implementation.
To verify this, we first provide the following result

\textbf{\textit{Lemma 1:}} \textit{For the wideband channel with $N_c$ taps, the {probability} that $k$ out of $K$ ($k\leq K$) beams reside within one tap is
approximated as
\begin{align}
P(K,k)=C_{K}^{k}\left(\frac{1}{N_c}\right)^k\left(\frac{N_c-1}{N_c}\right)^{K-k}.
\end{align}}

\noindent A brief illustration is made under $N_c=128$ and $K=10$. In this case, $P(10, 4)<10^{-5}$,
implying that it is virtually impossible for one tap containing over $3$ beams, so {$k$ is expected to be} smaller than 4,
regardless to say 10 for $P(10, 10)<10^{-18}$.
Combing above discussions, a proposition is made below to provide guidance on iterations setting:

\textbf{Proposition 2} [Number of iterations]: \textit{Let $P_T$ be a small threshold (e.g, $10^{-3}$) and $D$ be the number of effective taps after tap identification. A proper iterations can be set as $k-1$, where $k$ is the smallest integer satisfying $P(D,k)<P_T$.}

 To address P.2, DPC-BEM model was used in \cite{c8b} to capture the variations before implementing BOMP.
  This approach can dramatically lower the deterioration, but has two drawbacks.
  First, the estimation performance is heavily dependent on the basis order.
Secondly, to construct orthogonal DPC basis,
 a large-scale eigenvalue decomposition (EVD) has to be involved\cite{c17} with complexity $\mathcal{O}(L^3)$.
 To lower complexity while remaining robustness against Doppler,
{A-BOMP is proposed with its pseudo-code presented in \textbf{Algorithm 1}}. In A-BOMP, each outer iteration consists of three parts:

 \begin{algorithm}[t]
      \caption{Proposed A-BOMP Algorithm}
      \begin{algorithmic}
        \Require
         \textit{ Received signal $\bby_d$ and sensing matrix $\bm{\widetilde{\Psi}}_d$, maximum block-sparsity $\mathcal{K}$, group size $S$, group number $G=\frac{L}{S}$, and
          error threshold $\epsilon$.}
        \Ensure
         \textit{ The AoA support set $\mathcal{\widetilde{A}}_d$ and the corresponding AoD support set $\mathcal{\widetilde{D}}_d$.}
        \State \textit{Initialization: The residue $\bbr_d=\bby_d$, iteration index $C=0$, $\mathcal{A}_d$/$\mathcal{\widetilde{A}}_d$ and $\mathcal{D}_d$/$\mathcal{\widetilde{D}}_d$ are set to be empty, $\beta=\infty$, $\bm{\Phi}=\varnothing$ and $x=x_0=0$}.
        \While {$C<\mathcal{K}$ and $\beta>\epsilon$}
           \State $C=C+1$;
            \State $g_i=\mathop{\arg\max}\limits_{g}\sum\limits_{i=1}^{G}\frac{\|[\bm{\widetilde{\Psi}}^*_{d,g}\bbr_d]((i-1)S+1:iS)\|_1}{\|\bm{\widetilde{\Psi}}_{d,g}\|_F}$
          \State $n_R=\lceil g_i/G_t\rceil$ and $n_T=g_i-(n_R-1)G_t$.
          \If {$\exists~i, \mod(\mid \hspace{-0.1cm} n_{D}/n_R\hspace{-0.1cm}-\hspace{-0.1cm}\mathcal{D}(i)/\mathcal{A}(i)\hspace{-0.1cm}\mid,G_t/G_r)\leq 1$}
           \State \textbf{goto}~2
          \EndIf
           \State $\mathcal{A}=\{\mathcal{A},n_R\}$, $\mathcal{D}=\{\mathcal{D},n_T\}$
        \State $\widehat{\bbA}_T=[\bbf_{N_t}(\frac{n_T-1}{G_t}+\frac{2j_T}{G^2_t})]_{j_T\in[-\frac{G_t}{2}:1:\frac{G_t}{2}-1]}$
        \State $\widehat{\bbA}_R=[\bbf_N(\frac{n_R-1}{G_r}+\frac{2j_R}{G^2_R})]_{j_R\in[-\frac{G_r}{2}:1:\frac{G_r}{2}-1]}$
        \State $\bm{\widehat{\psi}}[n_i]=(\bbp^{'}_A[n_i]\otimes\bbw^*_A[n_i])(\widehat{\bbA}^*_T\otimes\widehat{\bbA}_R)_{{i=1\sim L}}$
        \State $\bm{\widehat{\Psi}}_{d,i}=\mathbf{diag}\big[\bm{\widehat{\psi}}[n_1](i),\cdots,\bm{\widehat{\psi}}[n_{G_tG_r}](i)\big]_{i=1\sim G_tG_r}$
        \State $\widehat{g}_i=\mathop{\arg\max}\limits_{g}\sum\limits_{i=1}^{G}\frac{\|[\bm{\widehat{\Psi}}^*_{d,g}\bbr_d]((i-1)S+1:mS)\|_1}{\|\bm{\widehat{\Psi}}_{d,g}\|_F}$
          \State $\widehat{n}_R=\lceil \widehat{g}_i/G_t\rceil$ and $\widehat{n}_T=\widehat{g}_i-(\widehat{n}_R-1)G_t$.
          \State $\bm\Phi\hspace{-0.1cm}=\hspace{-0.1cm}\big[\bm\Phi,\bbM({\bbf}^*_{N_t}(\frac{G_t(n_T\hspace{-0.05cm}-\hspace{-0.05cm}1)+\widehat{n}_T}{G^2_t})\hspace{-0.05cm}\otimes\hspace{-0.05cm}{\bbf}_{N_r}(\frac{G_r(n_R\hspace{-0.05cm}-\hspace{-0.05cm}1)+2\widehat{n}_R}{G^2_r})\big]$
          \For {$j=1:G$}
                \State $\bbj=(j-1)S+1:mS$
                \State $x=x+\|\bm\Phi(\bbj,:)^{\dagger}\bby(\bbj)\|_2$
                \State $\bbr_d(\bbj)=\bby_d(\bbj)-\bm\Phi(\bbj,:)\bm\Phi(\bbj,:)^{\dagger}\bby(\bbj)$
          \EndFor
          \State $\beta=|x-x_0|/x$
          \State $x_0=x$, $x=0$
          \State $\mathcal{\widetilde{A}}_d=\big\{\mathcal{\widetilde{A}}_d,\frac{2\pi G_r(n_R-1)+\widehat{n}_R}{G^2_r}\big\}$
          \State $\mathcal{\widetilde{D}}_d=\big\{\mathcal{\widetilde{D}}_d,\frac{2\pi G_t(n_T-1)+\widehat{n}_T}{G^2_t}\big\}$
        \EndWhile
      \end{algorithmic}
    \end{algorithm}
\begin{enumerate}[{$\mathbf{S}_1$}]
\item (Lines.5-9) \textit{partial basis matching}: select the angle pair having the largest sum of grouping correlations, and make sure that there is no overlapping with the selected ones.
\item (Lines.10-16) \textit{resolution refinement}: re-construct sensing matrix associated with the selected angle pair, and implement estimation procedure like $\mathbf{S}_1$ to refine resolution.
\item (Lines.17-21) \textit{partial residue update}: estimate the coefficients by the least-squared (LS) estimator, then update the residue $\bbr_d$ by subtracting the projection of each group.
\end{enumerate}

In A-BOMP, another key parameter is the group size $S$ {(equivalent to the group number $G$).
In the presence of Doppler,
the size of the non-zero support always exceeds the number of measurements,
 hence a reliable estimation cannot be guaranteed by BOMP even with infinite training frames.
  However, such uncertainty automatically vanishes in the absence of Doppler, if sufficient training frames
  are available. This is because to the non-zero support has a fixed size.}
 The great shortage of measurements forces us to ``shrink'' the non-zero support. To this end, we divide $\widetilde{\bbh}_{d,i}$ defined in Eq. (\ref{bvc}) into $S$ groups, and those entries belonging to one group are highly correlated thus being treated equally. Therefore, the group division essentially performs the signal compression, and $S$ is nothing but the coherent interval.

\textbf{Proposition~3:} [Determination of the group size] \textit{Let $\tau$ denote a high-correlation coefficient (e.g,~0.707). A proper group size can be set as the largest $S$ satisfying
$\cos(\omega_{max}N_cS)\leq\tau$ and $\omega_{max}N_cS\leq\pi/2$.}

\noindent Proposition 3 indicates that a smaller $\omega_{max}$ results in a larger $S$. For $\omega_{max}=0$, i.e., a time-invariant channel, A-BOMP degenerates to BOMP as {$G=L/S=1$}. Besides, one can readily verify that estimating ${\widetilde{\bbh}_d}$ via BOMP and estimating $\overline{\bbh}_d(n_i)$ via OMP are equivalent.  Compared to BOMP, A-BOMP only introduces a few small-scale matrix inversions, and
simulations show that such minimal computational cost will bring in a significantly improved accuracy.

Based on the output of A-BOMP, the steering matrices for tap-$d$ channel are estimated as
\begin{subequations}
\begin{align}
&\widetilde{\bbA}_{r,d}\hspace{-0.12cm}=\hspace{-0.12cm}\big[\bbf_{N_r}\big({\mathcal{\widetilde{A}}_{d}}(1)/{2\pi}\big),\cdots,\bbf_{N_r}\big({\mathcal{\widetilde{A}}_{d}}(c_d)/{2\pi}\big)\big] \\
&\widetilde{\bbA}_{t,d}\hspace{-0.12cm}=\hspace{-0.12cm}\big[\bbf_{N_t}\big({\mathcal{\widetilde{D}}_{d}}(1)/{2\pi}\big),\cdots,\bbf_{N_t}\big({\mathcal{\widetilde{D}}_{d}}(c_d)/{2\pi}\big)\big]
\end{align}
\end{subequations}
with $c_d=\mathbf{cal}(\mathcal{\widetilde{A}}_d)$. The approximate beamspace representation for tap-$d$ channel bear the form as
\begin{align}\label{com_form}
\widetilde{\bbH}_d(n)&=\widetilde{\bbA}_{r,d}\mathbf{diag}(\widetilde{\bbg}_d(n))\widetilde{\bbA}^*_{t,d}.
\end{align}
where $\widetilde{\bbg}_d(n)$  consists of unknown path gains.
Despite that both the path gain and angle support can be simultaneously obtained via OMP when estimating time-invariant channels,
for the more general time-varying channels, an additional stage is still necessary to estimate the path gain/Doppler.

 So far, we have completed the {second part
of the random-probing stage}. Summarizing, the main steps are listed as
\begin{itemize}
\item Stack the receive samples for each identified tap.
\item Transform the samples into a generic block-sparse form.
\item Determine the iterations and group size for A-BOMP.
\item Apply A-BOMP to estimate the angle support.
\end{itemize}

\section{Joint Estimation of Path Gain \& Doppler}
At the random-probing stage, the effective taps are identified with their angle support obtained as well.
In this section, we proceed to estimate the remaining unknown path gain/Doppler at the so-termed steering-probing stage.
\subsection{Steering probing design}
To accurately estimate path gains and Doppler shifts, {beamforming will be performed using the estimated angle supports
to improve the receive SNR.}
 For tap-$d$ channel,
 construct such a set $\mathcal{I}_d$ whose element $i$ is $(\mathcal{A}_d(i),\mathcal{D}_d(i))$.
  Because of the finite angular resolution and side-lode effects,
 different $\mathcal{I}_d$'s may share the same element, {thus we get their union as
 \begin{align}
 \mathcal{I}=\mathcal{I}_{d_1}\bigcup\mathcal{I}_{d_2}\bigcup\cdots\bigcup\mathcal{I}_{d_{D}}.
 \end{align}
Further, all AoAs and AoDs are individually extracted from $\mathcal{I}$ and captured by $\mathcal{I}_A$ and $\mathcal{I}_D$, respectively.}
To facilitate beamforming, only the discrete AoD indices need to be fed back.
Without causing ambiguity, we reset the time instant at the steering-probing stage,
{and making the following definition.}

{\noindent \textit{\textbf{Steering-probing vector}:
At the steering-probing stage, denote $\bbp^S_t(n)$ and $\bbp^S_t(n)$ to be the RF
vectors at time instant $n$. To improve receive SNR, $p^S_{p,t}(n)$ (the $p$-th element of $\bbp^S_t(n)$) and $p^S_{q,r}(n)$ (the $q$-th element of $\bbp^S_r(n)$)
are designed as\cite{c8b}
\begin{subequations}
\begin{align}
p^S_{p,t}(n)&=\frac{1}{\sqrt{N_t}}e^{jQ\big((p-1)\mathcal{I}_D(\widehat{n})\big)}, p\in[1,N_t]\\
p^S_{q,r}(n)&=\frac{1}{\sqrt{N_r}}e^{jQ\big((q-1)\mathcal{I}_A(\widehat{n})\big)}, q\in[1,N_r]
\end{align}
\end{subequations}
with $\widehat{n}=\mod\big(\big\lfloor {n}/{N_c}\big\rfloor,cal(\mathcal{I})\big)$.}}

\noindent The probing vectors repeat every $cal(\mathcal{I})$ subframes, making sure that each beam can be steered once in each polling.

\subsection{Path gain/Doppler estimation}
At the $i$-th polling, stacking all tap-$d$ related samples yields
\begin{align}\label{P_samples}
\bby_{d,i}=\big[y\big(d+n_{i,0}\big),\cdots,y\big(d+n_{i,|\mathcal{I}|-1}\big)\big]^{'}
\end{align}
where $n_{i,j}=N_cj+{\mathbf{cal}(\mathcal{I})}N_ci$, $\forall j\in[0,cal(\mathcal{I}))$.
{Using the compact beamspace representation obtained in Eq.~(\ref{com_form}), each sample in $\bby_{d,i}$ is approximately equivalent to}
\begin{align}\label{a10}
&y(n_{i,j}+d)\approx(\bbp^S_r(n_{i,j}))^*\widetilde{\bbA}_{r,d}\mathbf{diag}(\widetilde{\bbg}_d(n_{i.j}+d))\nonumber \\
&~~~~~~~~~~~~~~~~~~~~~~~~~\times\widetilde{\bbA}^*_{t,d}\bbp^S_t(n_{i,j})+\xi(n_{i,j}+d)\nonumber \\
&=\mathbf{vec}^{'}(\mathbf{diag}(\widetilde{\bbg}_d(n_{i,j}+d)))\bbm_d(n_{i,j})+\xi(n_{i,j}+d)
\end{align}
where $\bbm_d(n_{i,j})=\big((\bbp^S_t(n_{i,j}))^{'}\widetilde{\bbA}^*_{t,d}\big)\otimes\big((\bbp^S_r(n_{i,j}))^{*}\widetilde{\bbA}_{r,d}\big)$.
{By capturing the path gain with the one sampling in the middle of current polling,
 $\bby_{d,i}$ can be approximately represented as}
\begin{align}\label{pol}
\bby_{d,i}&\approx\underbrace{\left[\hspace{-0.15cm}
                  \begin{array}{c}
                    \bbm_d\big(n_{i,0}\big) \\
                    \bbm_d\big(n_{i,1}\big) \\
                    \vdots \\
                    \bbm_d\big(n_{i,cal(\mathcal{I})-1}\big) \\
                  \end{array}
                \hspace{-0.15cm}\right]}_{\bbM_{d,i}}\hspace{-0.05cm}\mathrm{vec}\bigg(\mathbf{diag}\big(\bbg_d(\overline{n}_i)\big)\bigg)\hspace{-0.05cm}+{\bm{\xi}_{d,i}}
\end{align}
with $\overline{n}_i\hspace{-0.1cm}=\hspace{-0.1cm}\big(n_{i,0}+n_{i,\mathbf{cal}(\mathcal{I})-1}\big)/2+d$ and ${\bm{\xi}_{d,i}}\hspace{-0.05cm}=\hspace{-0.05cm}[\xi(n_{i,0}+d),\xi(n_{i,1}+d),\cdots,\xi(n_{i,\mathbf{cal}(\mathcal{I})-1}+d)]^{'}$.
Let
$\widetilde{\bbM}_{d,i}\hspace{-0.05cm}=\hspace{-0.05cm}\big[\bbM_{d,i}[:,1^2],\bbM_{d,i}[:,2^2]\cdots,\bbM_{d,i}[:,C^2_d]\big]$, then Eq. (\ref{pol}) equals to
\begin{align}
\bby_{d,i}&\approx{\widetilde{\bbM}_{d,i}}\bbg_d(\overline{n}_i)+{\bm{\xi}_{d,i}}
\end{align}
Since $\mathbf{cal}(\mathcal{I})\geq c_d$, $\bbg_d(\overline{n}_i)$ can be recovered by LS estimator:
\begin{align}
\widehat{\bbg}_d(\overline{n}_i)=\widetilde{\bbM}^{\dagger}_{d,i}\bby_{d,i}=\bbg_d(\overline{n}_i)+\widetilde{\bbM}^{\dagger}{\bm{\xi}_{d,i}}.
\end{align}
 Once getting a new $\widehat{\bbg}_d$, we pick its $j$-th element, which is
 the estimated path gain of the $j$-th beam in current polling. The polling lasts for $R=\lfloor L/\mathbf{cal}(\mathcal{I})\rfloor$ times\footnote{
 {Similar to the random-probing stage, we introduce the  steering-probing state based on one frame
 consisting of $L$ subframes.
 In practice or numerical comparisons, one can simply replace $L$ with the actual
number of subframes, i.e., $\mathbf{cal}(I)R$.}},
so a pseudo time series is finally obtained as
\begin{align}
\widehat{\bbg}_{d,j}\hspace{-0.05cm}=\hspace{-0.05cm}\big[\widehat{g}_{d,j}(\overline{n}_0),\widehat{g}_{d,j}(\overline{n}_1),\cdots,\widehat{g}_{d,j}(\overline{n}_{R-1})\big]^{'}.
\end{align}

\textbf{\textit{Lemma 2:}} \textit{Through repetitive polling, the pseudo time series $\widehat{\bbg}_{d,j}$ has an equal sampling interval thus can be modeled as finite noisy samples of a single-tone sinusoid.}

Many techniques have been proposed over the years for the frequency estimation of a complex sinusoid in complex
additive white Gaussian noise. Here we adopt the WNALP estimator known for its computational efficiency and near-optimal performance\cite{c18}. The detailed procedures for path gain/Doppler estimation are described as below
\textit{\begin{itemize}
\item  Set $M_0=\lfloor R/2\rfloor$.
\item  Calculate the autocorrelation of $\widehat{\bbg}_{d,j}$ as
\begin{align}
R(m)=\frac{1}{R-m}\sum\limits_{i=m+1}^{2M_0}\widehat{g}_{d,j}(\overline{n}_i){\widehat{g}}^*_{d,j}(\overline{n}_{i-m})
\end{align}
\item Calculate the smoothing coefficient $w_m$ as
\begin{align}
w_m=\frac{3\big((M_0-m)(2M_0-m+1)-M^2_0\big)}{M_0(4M^2_0-1)}
\end{align}
\item Estimate the Doppler shift as
\begin{align}
\widehat{\omega}_{d,j}\hspace{-0.05cm}=\frac{1}{N_ccal(\mathcal{I})}\hspace{-0.1cm}\sum\limits_{m=1}^{M_0}w_m\mathbf{angle}(R(m)R^*(m\hspace{-0.05cm}-\hspace{-0.05cm}1))
\end{align}
\item Estimate the path gain as
\begin{align}
\widehat{g}_{d,j}(d)&=\frac{e^{-j\widehat{\omega}_{d,j}\frac{N_ccal(\mathcal{I})}{2}}}{R}\hspace{-0.1cm}\sum\limits_{i=1}^{R}\widehat{g}_{d,j}(\overline{n}_i)e^{-j\widehat{\omega}_{d,j}N_ccal(\mathcal{I})(i-1)}\nonumber\\
&=\frac{1}{R}\sum\limits_{i=1}^{R}\widehat{g}_{d,j}(\overline{n}_i)e^{-j\widehat{\omega}_{d,j}N_ccal(\mathcal{I})(i-1/2)}.
\end{align}
\end{itemize}}
\noindent The rest beams can be estimated similarly thus being omitted.

Summarizing, {the steering-probing stage is} carried out as:
\begin{enumerate}
\item Perform beamforming polling based on the union of the angle supports.
\item Estimate the path gains/Doppler for each tap via WNALP estimator.
\end{enumerate}

\section{Discussions and Simulations}

\subsection{Implementing Discussions}
\textit{Storage demand}: The major storage demand in channel estimation comes from the sensing matrix.
{In practice, suppose $p_1$ frames (the actual number of subframes is $p_1L$)} are allocated at the random-probing stage, then the sensing matrix size $C_1=p_1N\times UN_cG_tG_r$ in \cite{c10}, with
 $U$ being up-sampling ratio. Although our proposed estimation is conducted at each tap independently, the sensing matrix is shared by all taps with size $C_2=p_1L\times G_tG_r$. For $N_c=128$ and $N=512$, $C_1$ is more than over 13000 times larger than $C_2$.

\textit{Computational complexity}: The major computational complexity in channel estimation comes from the OMP-based algorithm, which comprises three steps in each outer iteration: basis matching, orthogonal projection, and residue update. For a sparse vector recovered via a {$V\times Q$} matrix, these steps {in the $k$-th iteration} require $(2V-1)Q$, $4kV$, and $2kV$ flops, respectively\cite{c19}.
The total flops of \cite{c10} and ours are $p_1NV(2UN_cG_tG_r+3p_1N)$ and $p_1LVG(2G_tG_r+3p_1L)$, both in the order of $\mathcal{O}(p_1^2)$.
Thanks to our extremely small-scale sensing matrix, even for $p_1<300$, the former is still more than ten times larger than the upper-bound of the latter.

\textit{Sensing matrix construction:} To ensure a reliable recovery via OMP, the sensing matrix should best satisfy the restricted isometry property (RIP). According to \cite{c20}, the optimal sensing matrix in terms of the RIP is the independent and identically distributed (IID) Gaussian matrix. Unfortunately, due to the constant-modulus limitation of the APS, the optimal sensing matrix cannot be realized and its design remains a open topic.
In this work, we follow {\cite{c8b,c9} and randomly draw the phase angle of each APS component from
the angle ensemble with an equal probability.}

{\textit{From a single RF chain to multiple RF chains:} Although the DS-DS channel estimator is introduced based on a single RF chain,
it can be readily generalized to multiple RF chains. Because the proposed estimator is relevant only to RF precoder only without a dedicated
digital precoding design throughout the entire estimation procedures similar to \cite{c12_1}.
 Assume $N_{RF}>1$ RF chains are employed at the transceiver.
 At the random-probing stage, each entry of the RF precoder is constructed following the rule described above,
 and the number of effective measurements increases by $N_{RF}$ times; at the steering-probing stage, the RF precoder
 sends $N_{RF}$ steering vectors each time, thus the estimated beams are steered $N_{RF}$ times as compared with the single-RF chain case.
  Besides this changes in the number of effective measurements,
  the algorithm that we described can be carried out without any modification.}
\begin{figure}[t]
\centering
    \centering
    \includegraphics[width=3.3in,height=3.2in]{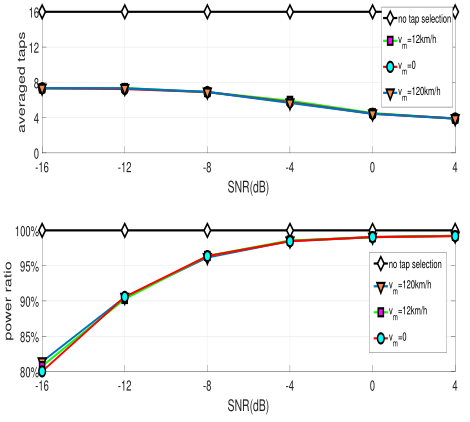}
    \caption{\small The averaged selected taps after tap identification}\label{Fig.5}
\end{figure}
\subsection{Simulation Verifications}

In this subsection, extensive numerical results are presented to verify the advantages of our proposed approach over existing works.
In simulations, the system carrier frequency $f_c$ is 60 GHz. The number of antennas is $N_t=N_r=32$, The dictionary sizes are $G_t=G_r=64$. $h(\cdot)$ is the raised-cosine filter with {the roll-off factor} $\beta=1$.  Each channel realization is generated according to Eq. (\ref{a1}) with $P$ ranging from 1 to 4. {If not specified,  the resolution of APS is 2-bit.}
 Other simulation parameters include $N_c=16$, $N=64$, $T_s=50ns$, $A=8$, $P_T=10^{-3}$, $\epsilon=0.01$ and $\mu=0.03$. The SNR (averaged TSNR) is defined as $\frac{L}{N\sigma^2}$. The estimation performance is weighted by the
normalized MSE (NMSE) given by
\begin{align}
\varepsilon=\frac{\sum_{d=0}^{N_c-1}\parallel\bbH_d-\widehat{\bbH}_d\parallel_F}{\sum_{d=0}^{N_c-1}\parallel\bbH_d\parallel_F}.
\end{align}

\subsubsection{Verification of the functionality of tap identification}
To verify the effectiveness of tap identification, we plot the averaged selected taps together with their power ratio in Fig. 3.
 $P=3$ and 40 frames are allocated at the random-probing stage. Three different {$v_{m}$'s: 0, 12km/h, and 120km/h} are considered.
We see that the tap identification
is regardless of {Doppler effects}. {Fewer taps are selected with the increase of SNR, and reduction is 75\% at 0dB.}
Note that, such a large reduction in processed taps is not at the cost of power loss.
As {can be} seen from Fig. 3, the averaged power ratio soon exceeds 97\% at medium SNR. The effectiveness of tap identification
is attributed to the delay-domain sparsity of mmWave channels.

\begin{figure}[t]
\centering
    \centering
    \includegraphics[width=3.3in,height=2.9in]{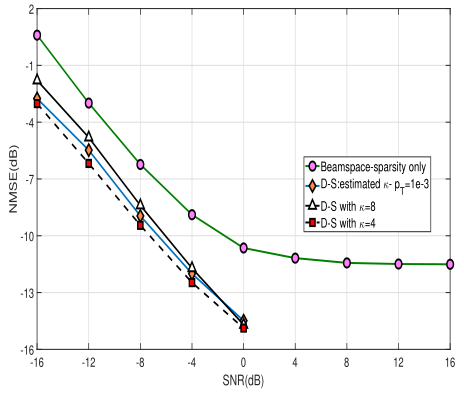}
    \caption{\small The NMSE comparisons among different schemes in static wideband channels}\label{Fig.5}
\end{figure}

\begin{figure}[t]
\centering
    \centering
    \includegraphics[width=3.3in,height=2.9in]{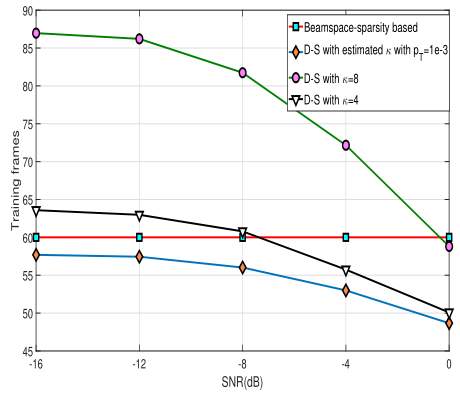}
    \caption{\small The total training frames consumed by different schemes in static wideband channels}\label{Fig.5}
\end{figure}

\begin{figure}[t]
\centering
    \centering
    \includegraphics[width=3.3in,height=2.9in]{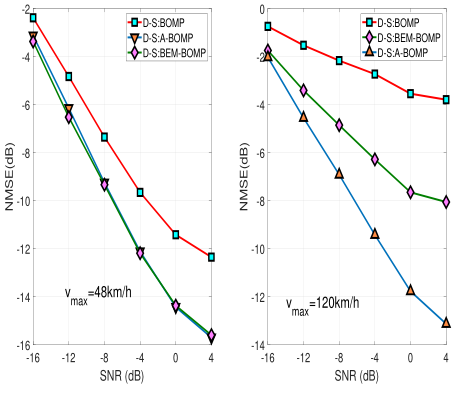}
    \caption{\small The NMSE comparisons in ``frequency-flat''\& time-varying channels in modest mobility}\label{Fig.5}
\end{figure}

\begin{figure}[t]
\centering
    \centering
    \includegraphics[width=3.3in,height=2.9in]{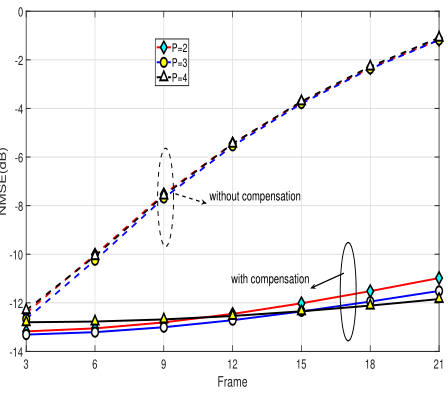}
    \caption{\small The NMSE versus the number of paths}\label{Fig.5}
\end{figure}

\begin{figure}[t]
\centering
    \centering
    \includegraphics[width=3.3in,height=2.9in]{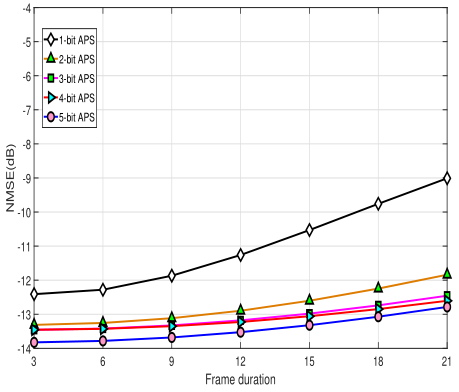}
    \caption{\small The NMSE versus the resolution of APS}\label{Fig.5}
\end{figure}

\subsubsection{NMSE comparisons in static \& wideband channels}
We then compare the double-sparse approach (DSA) with state-of-the-art  beamspace-sparse
approach (BSA)\cite{c10} at the same averaged SNR in Fig.~4. The channel is generated with 3 paths and $\omega_m=0$. For DSA, $40$ training frames are allocated at the random-probing stage with repeating beamforming polling for $R=4$ times at the steering-probing stage. 60 training frames are allocated for BSA and the regularized LS-estimator.
For BSA, its sensing matrix size is $16384\times 131072$, requiring a memory space over 18GB,
{in contrast to ours with size} $200\times 4096$ occupying 9Mb memory space. Due to the great shortage of training frames,
{the LS estimator without utilizing any sparsity performs the worst.
BSA performs much better than LS but still much worse than  DSA.} In addition,
 the shortage of probings makes the NMSE curve of BSA soon becomes flat. Even under the same peak SNR,
 we see that DSA still outperforms BSA at medium-to-high SNR region, {implying that the benefits brought by DSA outweight the
 power inefficiency of the proposed training pattern.}

In Fig.~5, we further plot the averaged consumed training frames of different approaches.
From two figures, it is clear that improper iterations ($\mathcal{K}=8$) will  {result in additional training overhead without making any }substantial performance improvement.
Following proposition 2, iterations can be properly set for A-BOMP (A-BOMP is equivalent to OMP here),
and the resultant NMSE performance is very close to the ideal benchmark ($\mathcal{K}=4$).
With pre-determined iterations, DSA requires the least training overhead, with a reduction of 20\% compared to BSBA at high SNR.
\subsubsection{NMSE comparisons in ``frequency-flat'' \& time-varying channels}
In Fig.~6, the channel is generated with $P=3$ with $v_m$=48km/h and $v_m$=120km/h, respectively.
We compare the NMSEs with the angle support recovered via A-BOMP and DPC-BOMP \cite{c8b}, respectively.
Since each tap channel is ``frequency-flat'' \& time-varying (FTV), the results are essentially the comparison with state-of-the-art
FTV channel estimator\cite{c8b}. $p_1=60$ frames are allocated at the random-probing stage with repeating beamforming polling $R=4$ times at the steering-probing stage. The DPC-basic order is $2$ as in \cite{c8b}.
In modest mobility ($v_m$=48km/h), A-BOMP and DPC-BOMP achieve similar performances, both outperforming BOMP remarkably.
In high mobility ($v_m$=120km/h), the advantage of A-BOMP over DPC-BOMP becomes notable.
As described before, A-BOMP avoids the large-scale EVD required in DPC-BOMP, demonstrating that it is more efficient and superior.

\subsubsection{NMSE performance in doubly-selective channels}
To thoroughly evaluate the functionality of DSA, we fix SNR=$-1$dB and $v_m$=$55$km/h,
and then simulate the NMSE versus the frame duration under various
conditions.

In Fig.~7, we compare the NMSEs by varying the number of paths. Other parameters are set as $p_1=60$, $R=4$, and $b=2$.
The results show that, without Doppler compensation,
the NMSE is soon to exceed $-10$dB, resulting in a great discrepancy with the actual channels.
By compensating for the Doppler using the estimate, superb tracking ability can be guaranteed over up to $20$ frames.
Furthermore, there is basically no NMSE degradation with the variation of $P$, indicating that DSA is not sensitive to the number of paths.

In Fig.~8, we compare the NMSEs by varying the resolution ($b$) of APS.
Other parameters are set as $p_1=60$, $b=2$, $R=4$, and $P=3$.
A remarkable performance gap is noticed with the ultra-coarse 1-bit APS. However, increasing $b$ by 1 bit will lead to a huge improvement.
The performance gap  compared to a finer APS (3$\sim$5-bit) in terms of the NMSE is very small (only 0.5dB), implying that
the proposed channel estimator is insensitive to the resolution of APS.

In Fig.~9, we compare the NMSEs by varying the number of RF chains. Other parameters are set as $P=3$, $b=2$, $R=6$, and $p_1=30$.
As can be seen, multiple RF chains can lower the estimation error compared to the single RF chain. This is because multiple RF chains
can generate more random beam probing patterns, which in turn benefits the recovery of the angle support using CS. We need to mention that
throughout the estimation, all non-zero symbols are set as one. Actually,
if the peak to average power ratio (PAPR) is not a significant concern, one can potentially set these symbols as the Gaussian distributed variables like\cite{c9} to
 strengthen the randomness.

\begin{figure}[t]
\centering
    \centering
    \includegraphics[width=3.3in,height=2.9in]{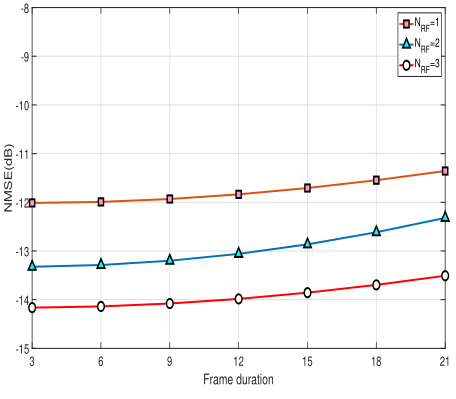}
    \caption{\small The NMSE versus the number of RF chains}\label{Fig.5}
\end{figure}
\section{Conclusion}
In this paper, we investigated the doubly-selective channel estimation for hybrid mmWave mMIMO systems.
Based on a judiciously designed training pattern, the beamspace sparsity and the delay-domain sparsity are jointly exploited to facilitate estimation. More importantly, the proposed channel estimator demonstrates strong abilities to combat the double selectivity
whilst leveraging the double sparsity.
Compared with existing works, our proposed doubly-sparse approach is demonstrated to be a more general and superior
solution to channel estimation under hybrid mmWave mMIMO.

\ifCLASSOPTIONcaptionsoff
  \newpage
\fi


\begin{thebibliography}{1}
\bibitem{0}
S. Gao, X. Cheng, and L. Yang,
``Making Wideband Channel Estimation Feasible for mmWave Massive MIMO: A Doubly Sparse
Approach,'' \emph{in Proc of IEEE International Conference on Communications.}, May, Shanghai, 2019

\bibitem{c1}
 F. Boccardi, R. W. Heath, A. Lozano, T. L. Marzetta, and P. Popovski,
``Five disruptive technology directions for 5G,'' \emph{IEEE Commun. Mag.},
vol. 52, no. 2, pp. 74-80, Feb. 2014.

\bibitem{c2}
 F. Rusek et al., ''Scaling up MIMO: Opportunities and challenges with
very large arrays,'' \emph{IEEE Signal Process. Mag.}, vol. 30, no. 1, pp. 40-60,
Jan. 2013.

\bibitem{c3}
T. S. Rappaport, J. N. Murdock, and F. Gutierrez, ``State of the art in 60-
GHz integrated circuits and systems for wireless communications,'' \emph{Proc.
IEEE,} vol. 99, no. 8, pp. 1390-1436, Aug. 2011.

\bibitem{c4}
O. El Ayach, S. Rajagopal, S. Abu-Surra, Z. Pi, and R. Heath, ``Spatially
sparse precoding in millimeter wave MIMO systems,'' \emph{IEEE Trans.
Wireless Commun.}, vol. 13, no. 3, pp. 1499-1513, Mar. 2014.




\bibitem{c6}
C. Chen, Y. Dong, X. Cheng, and L. Yang,
``Low-Resolution PSs Based Hybrid Precoding for Multi-User Communication Systems,'' \emph{IEEE Trans. Veh. Technol.}, vol. 67, no. 7, pp. 6037-
6047, July 2018.

\bibitem{c7}
S. Gao, Y. Dong, C. Chen, and Y. Jin, ``Hierarchical beam selection in mmWave multiuser MIMO systems with one-bit analog phase shifters,''
in \emph{Proc. IEEE WCSP}, Yangzhou, China, Oct. 2016.


\bibitem{c7a}
S. Gao, X. Cheng, and L. Yang,
``Spatial Multiplexing With Limited RF Chains: Generalized Beamspace Modulation (GBM) for mmWave Massive MIMO,'' \emph{IEEE J. Sel. Areas Commun.}, vol. 37, no. 9,
pp. 2029-2039, Sep. 2019.

\bibitem{c7b}
F. Rusek, D. Persson, B. K. Lau, E. G. Larsson, T. L. Marzetta, O.
Edfors, and F. Tufvesson, ``Scaling up MIMO: Opportunities and challenges with very large arrays,'' \emph{IEEE Signal Process. Mag.,} vol. 30, no.
1, pp. 40-46, Jan. 2013.


\bibitem{c7c}
S. Gao, X. Cheng, L. Yang and R. Zhang, ``Zero-Forcing Based Limited Feedback Hybrid Precoding in mmWave Communications,'' in \emph{Proc. IEEE International Conference on Communications in China}, Changchun, China, Aug. 2019.

\bibitem{c8}
A. Alkhateeb, O. El Ayach, G. Leus, and R. Heath, ``Channel estimation
and hybrid precoding for millimeter wave cellular systems,'' \emph{IEEE J. Sel.
Topics Signal Process.}, vol. 8, no. 5, pp. 831-846, Oct. 2014.


\bibitem{c8a}
 J. A. Tropp and A. C. Gilbert, ``Signal recovery from random measurements via orthogonal matching pursuit,'' \emph{IEEE Trans. Inf. Theory},
vol. 53, pp. 4655-4666, Dec. 2007.

\bibitem{c8b}
Q. Qin, L. Gui, and P. Cheng,
``Time-Varying Channel Estimation for Millimeter Wave Multiuser MIMO Systems,'' \emph{IEEE Trans. Veh. Technol.}, vol. 67, no. 10, pp. 9435-
9448, Oct 2018.

\bibitem{c9}
Z. Gao, C. Hu, L. Dai, and Z. Wang, ``Channel estimation for millimeterwave
massive MIMO with hybrid precoding over frequency-selective
fading channels,'' \emph{IEEE Commun. Lett.}, vol. 20, no. 6, pp. 1259-1262,
Jun. 2016

\bibitem{c10}
K. Venugopal, A. Alkhateeb, N. Gonzalez-Prelcic, and R. Heath,
``Channel estimation for hybrid architecture-based wideband millimeter
wave systems,'' \emph{IEEE J. Sel. Areas Commun.}, vol. 35, no. 9,
pp. 1996-2009, Sep. 2017.

\bibitem{c11}
J. Fernandez, N. Prelcic, K. Venugopal, and R. Heath, ``Frequency-domain Compressive Channel Estimation for
Frequency-Selective Hybrid mmWave MIMO Systems,'' \emph{IEEE Trans. Wireless Commun,} vol. 17, no. 5,
pp. 2946-2960, May. 2018.


\bibitem{c11a}
C. Carbonelli, S. Vedantam, and U. Mitra, ``Sparse channel estimation
with zero tap detection,'' \emph{IEEE Trans. Wireless Commun.}, vol. 6, no. 5,
pp. 1743-1753, May 2007.

\bibitem{c11b}
B. Wang, F. Gao, S. Jin, H. Lin, and G. Y. Li, ``Spatial- and frequencywideband
effects in millimeter-wave massive MIMO systems,'' \emph{IEEE
Trans. Signal Process.}, vol. 66, no. 13, pp. 3393-3406, Jul. 2018


\bibitem{c11c}
J. Mo,  P. Schniter, and R. Heath, ``Channel Estimation in Broadband Millimeter Wave
MIMO Systems With Few-Bit ADCs,'' \emph{IEEE
Trans. Signal Process.}, vol. 66, no. 5, pp. 1141-1154, Mar. 2018

\bibitem{c12}
A. Alkhateeb and R. Heath, ``Frequency selective hybrid precoding
for limited feedback millimeter wave systems,'' \emph{IEEE Trans.
Commun.,} vol. 64, no. 5, pp. 1801-1818, May 2016.


\bibitem{c12_1}
K. Venugopal, A. Alkhateeb, R. W. Heath, Jr., and N. Gonz¨¢lez-Prelcic,
``Time-domain channel estimation for wideband millimeter wave systems
with hybrid architecture,'' \emph{in Proc. IEEE Int. Conf. Acoust., Speech
Signal Process. (ICASSP)}, Mar. 2017, pp. 6493-6497.

\bibitem{c12a}
R. Heath, ``Vehicular mmWave Communication: Opportunities and Challenges,'' Available:https://
users.ece.utexas.edu/rheath/presentations/2015/\\
VehicularMmWaveCommunicationOpportunitiesChallenges.pdf

\bibitem{c12b}
S. Buzzi and C. D'Andrea, ``On clustered statistical MIMO millimeter
wave channel simulation'' Available:https://
arxiv.org/abs/1604.00648, May 2016.

\bibitem{c13}
\emph{IEEE 802.11n} Wireless Lan Medium Access Control (MAC) and Physical Layer (PHY)
Specifications, \emph{IEEE Std 802.11-2012} (Revision of IEEE Std 802.11-
2007), 2012.

\bibitem{c13b}
X. Ma, L. Yang, and G. B. Giannakis, ''Optimal training for MIMO
frequency-selective fading channels,'' \emph{IEEE Trans. Wireless Commun.},
vol. 4, no. 2, pp. 453-466, Mar. 2005.


\bibitem{c13c}
C. Gustafson, K. Haneda, S. Wyne, and F. Tufvesson, ``On mm-wave
multipath clustering and channel modeling,'' \emph{IEEE Trans.
Antennas and Propag,} vol. 62, no. 3, pp. 1445-1455, March 2014.





\bibitem{c16}
\emph{IEEE 802.15} WPAN Millimeter Wave Alternative PHY Task Group
3C (TG3c). [Online]. Available: http://www.ieee802.org/15/pub/TG3c
contributions.html


\bibitem{c16a}
 X. He, R. Song, and W. P. Zhu, ``Pilot allocation for distributed
compressed sensing based sparse channel estimation in MIMO OFDM
systems,'' \emph{IEEE Trans. on Veh. Technol.}, vol. 65, pp. 2990-3004, May
2016.

\bibitem{c16b}
L. Liu, E. G. Larsson, W. Yu, P. Popovski, C. Stefanovic, and E. de Carvalho, ``Sparse signal processing for grant-free massive connectivity: A
future paradigm for random access protocols in the internet of things,''
\emph{IEEE Signal Process. Mag.}, vol. 35, no. 5, pp. 88-99, Sep. 2018.


\bibitem{c17}
T. Zemen and C. F. Mecklenbrauker, ``Time-variant channel estimation
using discrete prolate spheroidal sequences,'' \emph{IEEE Trans. Signal Process.,} vol. 53, pp. 3597-3607, Sep. 2005.


\bibitem{c18}
A. B. Awoseyila, C. Kasparis, and B. G. Evans, ``Improved single
frequency estimation with wide acquisition range,''\emph{Electron. Lett.}, vol.
44, no. 3, pp. 245-247, Jan. 2008.

\bibitem{c19}
J. Wang, S. Kwon, and B. Shim, ``Generalized orthogonal matching
pursuit,'' \emph{IEEE Trans. Signal Process.,} vol. 60, no. 12, pp. 6202-6216,
Dec. 2012.


\bibitem{c20}
M. A. Davenport and M. B. Wakin, ``Analysis of orthogonal matching
pursuit using the restricted isometry property,'' \emph{IEEE Trans. Inf. Theory,}
vol. 56, no. 9, pp. 4395-4401, Sep. 2010.


%

\end{thebibliography}
\end{document}